# Exploring the Regulatory Function of the N-terminal Domain of SARS-CoV-2 Spike Protein Through Molecular Dynamics Simulation

**Authors:** Yao Li[1,2,3†], Tong Wang[3†*], Juanrong Zhang[1,2,3†], Bin Shao[3], Haipeng Gong[1,2*], Yusong Wang[3], Siyuan Liu[3,4,5] and Tie-Yan Liu[3]

**Affiliations:**

[1]MOE Key Laboratory of Bioinformatics, School of Life Sciences, Tsinghua University, Beijing 100084, China
[2]Beijing Advanced Innovation Center for Structural Biology, Tsinghua University, Beijing 100084, China

[3]Microsoft Research Asia, Beijing, China

[4]School of Data and Computer Science, Sun Yat-sen University, Guangzhou, China

[5]Guangdong Key Laboratory of Big Data Analysis and Processing, Guangzhou, China

[†]These authors contributed equally.

[*]To whom correspondence should be addressed: watong@microsoft.com (T.W.) and hgong@mail.tsinghua.edu.cn (H.G.)

## Abstract

SARS-CoV-2 is what has caused the COVID-19 pandemic. Early viral infection is mediated by the SARS-CoV-2 homo-trimeric Spike (S) protein with its receptor binding domains (RBDs) in the receptor-accessible state. We performed molecular dynamics simulation on the S protein with a focus on the function of its N-terminal domains (NTDs). Our study reveals that the NTD acts as a “wedge” and plays a crucial regulatory role in the conformational changes of the S protein. The complete RBD structural transition is allowed only when the neighboring NTD that typically prohibits the RBD’s movements as a wedge detaches and swings away. Based on this NTD “wedge” model, we propose that the NTD-RBD interface should be a potential drug target.

## Introduction

The Coronavirus Disease 2019 (COVID-19), which has become a worldwide pandemic, is caused by the Severe Acute Respiratory Syndrome Coronavirus 2 (SARS-CoV-2) (1-3). Similar to other β coronaviruses such as SARS-CoV and MERS-CoV, SARS-CoV-2 is an enveloped, positive-stranded RNA virus (4, 5). Furthermore, SARS-CoV-2 has high homology with SARS-CoV, sharing the same human cell receptor, angiotensin-converting enzyme 2 (ACE2) (6, 7).

After the completion of the genome sequencing of SARS-CoV-2 (1), a great amount of research attention has been put on elucidating the infection process and searching for potential therapies against COVID-19. The experimentally determined structures of the Spike (S) protein (8, 9), Main protease (10), RNA dependent polymerase (11), as well as the complex formed by the S protein and ACE2 (12, 13) provide atomic details of SARS-CoV-2. Meanwhile, computational efforts have been exerted to improve the understanding of the infection mechanism via evolutional analysis (14, 15). According to previous studies, the first step of the infection is the attachment of the S protein to ACE2 (16). The homo-trimeric S protein is located on the surface of the viral membrane, with each monomer consisting of two regions, S1 and S2, which are respectively responsible for receptor binding and membrane fusion. S1 is composed of a receptor binding domain (RBD) and an N-terminal domain (NTD) (17). Reported infection models for other coronaviruses suggest that the association of ACE2 and S trimers triggers a series of molecular events, including the shedding of S1, refolding of S2, and subsequent fusion of viral and host cell membranes (18-21). Multiple conformations of the SARS-CoV-2 S protein have been identified, with varying orientations of the RBDs and distinct ACE2 binding affinities (*8*, *9*). The closed state with all three RBDs in the downward orientation is ACE2-inaccessible and thus inactive, while the partially open state with one RBD flipped up is capable of receptor binding (*9*). Recently, the semi-open state with two RBDs turned upwards as well as the open state with all three RBDs turned upwards have been detected by cryo-electron microscopy (Cryo-EM) (*22*), although the corresponding high-resolution structures have not been released yet.

With more atomic details of the S protein unveiled, the functional role of the RBD is generally understood by the research community. However, the contribution of the NTD, another component of S1, in the infection process is still elusive, despite some subtle clues. In MERS-CoV, a neutralizing monoclonal antibody 7D10 targeted at an NTD epitope precludes the RBD from binding to its cell receptor DPP4 by steric hindrance (*23*). Interestingly, in SARS-CoV-2, a monoclonal antibody 4A8 that directly interacts with the NTD without blocking the RBD-ACE2 binding still exhibits neutralizing effects (*24*). Recently, an obvious movement of the NTD was detected accompanying the RBD’s conformational changes in the D614G variant of the S protein (*22*). These studies suggest that the NTD is at least not completely independent of the functional conformational change of the RBD in the S protein and thus may be involved in the regulation of viral infection.

Molecular dynamics (MD) simulation is a computational tool for understanding the dynamics and function of macromolecules. Particularly, it can supplement the static spatial information provided by experimental methods such as X-ray crystallography and Cryo-EM. In our work, we performed MD simulations on four systems that corresponded to the closed, partially open, semi-open, and open states of the S protein. To explore the functional role of the NTD, each system was simulated 1 $\mu s$ using accelerated MD (aMD) (*25*) with 5 repeats to thoroughly sample the relative motion between the NTD and the RBD. Our results suggest that the upward RBD is intrinsically inclined to tilt towards to the more stable downward orientation, and that the NTD either favorably interacts with the RBD as a wedge to block this motion or occasionally detaches from the RBD to allow its free transition between the upward and downward orientations. Based on this "wedge" effect of the NTD, we virtually screened potential drugs that strengthen the NTD-RBD interface, which would hypothetically prohibit the structural transition of RBD that is required for viral infection.

## Results

### Conformational changes of the S proteins with different initial structures

To study the conformational changes of the S proteins of SARS-CoV-2, we built four simulation systems of the S trimer with different initial conformations, i.e., a closed system with three "downward" RBDs, a partially open system with one "upward" RBD, a semi-open system with two "upward" RBDs, and an open system with three "upward" RBDs. The S proteins in the closed system and the partially open system were sourced from PDB structures 6VXX (*9*) and 6VYB (*9*), respectively, where non-native mutations in 6VYB were mutated back to the wild-type counterparts, and the missing fragments were reconstructed by SWISS-MODEL (*26*-*30*). Meanwhile, the initial conformations in the semi-open and open systems were built by homology modeling using SWISS-MODEL by taking the S proteins of SARS-CoV with 2 upward RBDs (PDB ID: 6NB6) (*21*) and 3 upward RBDs (PDB ID: 6NB7) (*21*) as templates, respectively. With an explicit water solvent model, each simulation system consists of nearly one million atoms. Considering the intimidating sizes of simulation systems and the simulation steps required for large-scale conformational changes, we initiated 5 independent one-microsecond aMD simulations (*25*) following 10-ns equilibration with a timestep of 4 fs (*31*) to enhance the conformational sampling in each system.

As shown in Fig. 1, remarkable conformational changes are observed in the S1 region, including both NTD and RBD, while the S2 region retains structural rigidity. Hence, we used the S2 region as a basis to define the central axis of the S protein (red dashed lines shown in Fig.1) and used the upper part of it as a reference for evaluating the conformational changes of the S1 region (see Methods for more details). The simulation results show that NTDs may play an important role in regulating the movement of the adjacent RBDs. In the closed system, besides the trend of one NTD moving away from the central axis observed in a trajectory, all RBDs were locked in the "downward" conformation, and the other NTDs were stable in all 5 repeats (Fig. 1A), in agreement with the experimental evidence that the closed S protein is stable and forms a condensed structure with RBDs packed tightly against the adjacent NTDs (*8, 19*-*21, 32*). In contrast, dramatic relative movements between NTDs and RBDs were observed in the other three simulation systems. In the partially open system, the upward RBD of chain B tended to tilt downwards while the adjacent NTD of chain C moved toward the central axis to prevent this downward movement; this observation was confirmed by all five repeated simulations (Fig. 1B). This blockage of RBD by NTD was also observed in the semi-open and open systems (Fig. 1C&D). Interestingly, an unusual event was detected in these two systems: when the NTD occasionally moved away from the central axis, the adjacent RBD was allowed to tilt downwards completely (Fig. 1E&F).

We proposed novel angle and distance metrics to quantify the movements of the NTD and RBD in the simulation trajectories. Previous research used three single residues, i.e. D405-V622-V991, to define the angle for evaluating the orientation of the RBD (*33*). To ensure robustness of the results, we replaced these single residues with the centroids of structurally rigid regions and defined three metrics: 1) RBD angle ($\theta_r$) that describes the upward and downward tilting poses of the RBD and any conformations in between; 2) RBD distance ($d_r$) that describes the distance between the RBD and the central axis; 3) NTD distance ($d_n$) that describes the distance between the NTD and the central axis (Fig. 2A, see Methods for more details). As shown in Fig. 2B-D, the quantitative measurements of the relative motions between NTDs and neighboring RBDs agree with the representative conformational changes shown in Fig. 1. All motions can be classified into two modes. In the first mode, the NTD wedges in to block RBD movement, as detected in all three simulation systems possessing upward RBDs. As shown in the first three columns of Fig. 2B-D, both angle and distance of the upward RBD decrease, but this tendency is later blocked by the quick movement of the NTD towards the central axis (i.e., a marked reduction of the NTD distance). In the second mode, NTD swings away, and RBD accomplishes its complete downward tilt, as observed in both the semi-open and open systems. As shown in the last two columns of Fig. 2D, the distance of the NTD increases in the early stage of simulation, suggesting that it moves away from the central axis, releasing the steric hindrance for RBD movement. Accordingly, the highlighted upward RBDs in both semi-open system (the penultimate column) and open system (the last column) eventually approach much smaller distance and angle values compared with the blocked RBDs (the first three columns), which indicates that these highlighted RBDs tilt downward completely. To further validate these two modes, we calculated the distributions of RBD distance and angles from all five repeats for each simulation system. Consistently, the metrics of the upward and downward RBDs are separated by a marked gap in the RBD blocked mode (see the first three columns of Fig. 2E&F), while huge overlaps occur in the RBD downward tilting mode (see the last two columns of Fig. 2E&F). Moreover, one-way ANOVA analysis confirms that significant difference is present between experimental groups and control groups in the RBD blocked mode but is absent in the RBD downward tilting mode (Table S1). These results imply that the NTD may act as a wedge to regulate the functional movement of the RBD.

The complete downward tilt of the RBD was not caught in the partially open system, possibly because the NTD kept wedging in during the simulations. To validate this hypothesis, we artificially forced the NTD of chain C to the swinging-out posture (using targeted MD (tMD) on a target position inferred from a reference structure with large NTD distance in the semi-open system) and then investigated the movement of the upward RBD in this system through 5 redundant 400-ns aMD simulations (Fig. S1A). Compared with previous simulations that started from the crystal structure, the distributions of RBD angles and distances exhibited remarkable left-shifts in the NTD-forced-out trajectories (Fig. S1B), indicating that the upward RBD tilts downward to a significantly larger extent when its partner NTD is absent. We also conducted a time-structure based Independent Components Analysis (tICA) (*34*, *35*) to identify slow motions with high time autocorrelation on the first 200 ns of these NTD-forced-out simulations. As expected, the first tIC presents the highest correlation with the difference vector between the representative structures of the upward and downward RBDs (Fig. S1C). Hence, the downward tilting movement of the RBD is the main slow motion when the NTD is forced away.

## Intermediate conformations and transition analysis by Markov State Models

Markov state models (MSM) (*36*) are a powerful tool used to extract structural information from multiple simulations and has been exploited to track dynamic progress and elucidate slow motions in MD simulations (*37-39*). In our work, considering the relative structural rigidity of the S2 region, we extracted Cartesian coordinates of Cα atoms of S1 to build MSM and identified intermediate conformational states in the two modes described previously. For the NTD wedging-in mode, all five repeated trajectories of the partially open system were used in the principal component analysis (PCA) to reduce the dimensionality of original coordinates. As shown in Fig. S2A, the top 4 principal components (PCs) were picked to represent the S1 regions of three chains since they explained 69.38% of the total variance. A lag-time of 1.0 ns was set for the time-discretization of the process (*39*, *40*) and a number of 100 microstates was chosen based on the implied timescale analysis (*37*, *41*). These microstates were built using mini-batch K-Means clustering (*40*, *42*, *43*) and were subsequently aggregated into a number of macrostates (or metastable states), for which pairwise transition probabilities were calculated (*34*). The detailed workflow of MSM construction is illustrated in Methods.

As shown in Fig. 3A&B, all five trajectories in the partially open system were finally lumped into six macrostates and the structure closest to the centroid of each macrostate cluster in the space of the top 4 PCs was chosen as the representative structure. State 0 is the initial state with one RBD turned upwards, whereas states 4 and 5 are the final states with this RBD blocked by the adjacent NTD. States 1, 2 and 3 are intermediate states between the initial and final states. With the NTD quickly moving toward the central axis in the early stage of simulation, the initial state 0 can directly transform into the final state 4 with a high probability. Meanwhile, in the intermediate state 2, the NTD initially moves a little away from the central axis to allow the upward RBD to tilt slightly downward, leading to significantly smaller RBD angles than those of the initial state (Fig. 3B). The NTD then moves back toward the central axis, preventing the upward RBD from falling, and consequently, the RBD angle increases afterwards, reaching a stable final state (state 5) with RBD held in the upward pose (Fig. 3B). State 3 is another intermediate state, which either transforms into the final state 5 with the NTD moving towards the central axis or interconverts with intermediate state 2.

We also built an MSM for the mode characterized by the NTD moving out and the RBD tilting downward from two trajectories of the semi-open system. Following the same protocol, we chose the top 5 PCs, set the microstate number as 100 and set the lag-time as 1.0 ns (Fig. S2B). As shown in Fig. 3C, this mode is represented by a four-state transition network, including two transition paths between the RBD's initial upward state (state 0) and the RBD's final downward state (state 3), mediated by two intermediate states (states 1 and 2). In this mode, the NTD first moves away from the central axis, eliciting the upward RBD to fall (state 1) and/or moves inward to stabilize the already downward RBD (state 2). As shown in Fig. 3D, the final state (state 3) exhibits the smallest values in all three metrics, implying that the upward RBD has tilted down completely to form a highly packed structure in this state. Notably, the transition probabilities here are obtained from aMD simulations with boosting potentials (*44*) and thus may not correlate well with the "real" probabilities estimated from an equilibrium distribution (*37*).

## Analysis on the interface between NTD and RBD during structural transition

As demonstrated above, each state in MSM is crucial to the transition from the initial to the final state, and the whole network clearly illustrates how the overall conformational change happens step by step. In order to figure out the molecular mechanism of these conformational changes, we performed Molecular Mechanics Generalized Born Surface Area (MM/GBSA) (*45-47*) on each microstate with one protomer as the ligand and the remaining two as the receptor and decomposed the total binding free energy to individual residue pairs. By this means, the residue pairs that make significant contribution to the physical interactions between the NTDs and their adjacent RBD partners can be identified.

The identified residue pairs are mainly located at the NTD-RBD interface but can be clustered spatially into several clumps (Table S2&S3). The interaction patterns of these clumps vary substantially during the structural transition. Specifically, for the mode characterized by NTD wedging in and RBD being blocked in the partially open system (Fig. 4A and Table S2), the red clump, including H519 and P230, forms in the intermediate states 2 and 3 and remains in the two final states 4 and 5. Meanwhile, the yellow clump, including P521 and Y170, forms and breaks intermittently and disappears in the final state 5. We suspect that the interactions between these residue pairs are more relevant in mediating the relative motions between the NTD and the RBD. On the contrary, the more conservative clumps that exist among most macrostates, e.g., the green clump including R357 and N165, as well as the purple clump including N360 and T167, are likely to play more basic roles such as structural stabilization. For the mode characterized by NTD moving out and RBD tilting downward in the semi-open system (Fig. 4B and Table S3), the identified residue pairs scarcely overlap among macrostates, possibly due to the more drastic relative motions between the NTD and the RBD. However, it is noteworthy that the initial state lacks strongly interacting residue pairs if -1.5 kcal/mol is taken as the energy threshold. Interactions gradually form when the RBD tilts downward, reaching the final stable state with the NTD-RBD interface composed of three clumps.

We further performed evolutional analysis on key residue pairs identified in the partially open system: the residues of the RBD of chain B from 355 to 360 and from 519 to 521, and the residues of the NTD of chain C from 159 to 170 and from 230 to 233. We first compared such residue pairs with the corresponding residues in SARS-CoV and MERS-CoV. As shown in Fig. 4C, most of these key residues are conserved in SARS-CoV while only parts of them are conserved among three kinds of sequences (conservativeness reflected by the darkness of blue colors), which is consistent with the overall sequence similarity among the three coronaviruses. When evaluated on the samples of COVID-19 patients derived from Global Initiative on Sharing All Influenza Data (GISAID) (*51*, *52*), all these residues were highly conserved regardless of region or race (Fig. 4D). These results indicate that the NTD-regulated structural transition may be generalizable for all current SARS-CoV-2 samples, which implies that potential therapy based on this mechanism may be applicable to all COVID-19 cases.

## The NTD wedge effect for conformational changes of the S protein

Given all the consistent observations and comprehensive analysis demonstrated above, we propose the "wedge" model to illustrate the crucial regulatory role of the NTD in the functional conformational change of the S protein. As shown in Fig. 5, the NTD acts as a wedge in two aspects: wedging in for RBD blockage and moving out to allow RBD tilting. In general, the receptor-accessible upward RBD is inclined to tilt downward to form the more compact structure. Occasionally, the NTD may swing away and detach from the RBD, sterically allowing the free movement of the RBD. Although not observed in our simulations due to the high stability of the closed S protein, the opposite motion of the RBD tilting upward may occur during the activation of the S protein. The free reorientation of the RBD, however, will be blocked when the neighboring NTD moves towards the central axis and forms favorable interactions with the RBD. According to this model, the NTD may be a potential drug target for COVID-19 therapy. Particularly, drugs targeting the NTD-RBD interface is supposed to augment the NTD "wedge" effect and lock the S protein into an individual conformational state (complete "up" or complete "down" states), which would inhibit the functional structural transition required for viral infection.

### Preliminary NTD-targeted drug virtual screening

Given the "wedge" model of the NTD, the partially open state was employed to recruit drug compounds that enhance interactions on the NTD-RBD interface. The NTD of chain C from the last frame in the equilibrium simulation of the partially open system was chosen as the receptor. We used AutoDock Vina (*55*, *56*) to perform virtual screening against the sub dataset of "in-trials" in ZINC15(*57-59*), including all approved drugs and investigational compounds in clinical trials. 9,800 chemical compounds were evaluated, and top compounds with high binding affinities are shown in Table S4. The compound ZINC000261494659, a Paclitaxel analogue, has the highest binding affinity of -13.8kcal/mol and binds at the interface between NTD of chain C and the "up" RBD of chain B. The drug compound strengthens the NTD-RBD binding by forming new interactions at the interface, including hydrogen bonds (e.g., with N360 of the RBD of chain B and V171 of the NTD of chain C), hydrophobic interactions (e.g., with P330 of the RBD and V159 of the NTD) and van der Waals (Fig. 6A). ZINC000256109538 can also bind at the NTD-RBD interface with high affinity but in a different pose (Fig. 6B), forming interactions with T108, Q115, and T167 of the NTD as well as I233, R357, and N394 of the RBD. In addition, we also performed virtual screening against the database of TCMSP (*60*), which consists of 14,249 chemical compounds extracted from Chinese herbs, and used a threshold of -9.5 kcal/mol to select compounds with high binding affinities to the NTD-RBD interface. Interestingly, the identified compounds are ingredients of the prescriptions recommended by the National Health Commission of China for the treatment of COVID-19 (Table S5-13). For example, in the prescription of "Qing Fei Pai Du Tang", a total of 18 compounds from 8 herbs were detected with high binding affinities (Fig. 6D). These results imply that active constituents of the recommended prescriptions may deliver their medical effects by targeting the NTD of the S protein.

## Discussion

In our work, we found that the NTD acts as a wedge to regulate the movement of the adjacent upward RBD: holding it to stand up (the "wedging in" mode) or allowing it to tilt down (the "moving out" mode). Both modes were observed in the simulation systems with upward RBDs, and several metastable states were identified by MSM. This NTD "wedge" model should be further validated by biochemistry and molecular biology experiments. Notably, previous research has proven that the closed and the partially open structures coexist *in vivo* (*61*), which implies active transitions between the two structure states. However, despite the energy boost applied, we failed to observe significant conformational changes of RBDs in the simulations of the closed system. More advanced enhanced sampling techniques or even biased simulations may be conducted in the future to see the RBD's upward flipping movement. Alternatively, based on the RBD's downward tilting trajectories observed in this work, a reaction path connecting the two conformational states could be identified and optimized by path optimization approaches like the string method, which allows thorough sampling and rigorous free energy estimation along the reaction coordinates.

The semi-open and open states are more flexible with more than one upward RBD, and thus their simulations exhibit high diversities. As shown in Fig. S3 and Fig. S4, besides the two kinds of relative movements between the NTD and the RBD, we also observed that the upward RBDs formed interactions with each other. In the semi-open system, two upward RBDs tilted down, and the three downward RBDs in the final structure formed stable mutual interactions (Fig. S3). Similar behavior was also observed in the open system (Fig. S4), where all three upward RBDs tilted downwards completely to form a compact packing. Occasionally, two RBDs tilted downwards while the remaining upward RBD only formed close interactions with one of the other two RBDs (Fig. S4B). Nevertheless, the complicated interactions with multiple upward RBDs and their potential influence on viral infection still await further investigation.

Our main observations were obtained from aMD simulations, which disrupted the Boltzmann distribution by applying energy boosts. Hence, unperturbed distributions of conformational states need to be recovered from the microstate transition analysis in MSM, which are essential for the evaluation of kinetic properties, such as transition probabilities, and thermodynamic properties, such as the potential of mean force. In addition, although we have screened out dozens of promising compounds with high binding affinities to the NTD-RBD interface from the ZINC15 and TCMSP databases, more stringent calculations and experimental validations should be conducted to evaluate the identified drug candidates in the future.

## Materials and Methods

### Simulation system design and configuration

The initial structures of the closed system and the partially open system were chosen from the PDB structures of 6VXX (*9*) and 6VYB (*9*), respectively. These structures are trimer glycoproteins with 1281 residues with N-glycosylation in each chain. In our research work, non-native proline mutations were mutated back to the wild type lysine and valine, respectively. Since several fragments were not resolved in these Cryo-EM structures, SWISS-MODEL (*26*-*30*) was employed to replenish the missing atoms. We used the GLYCAM-Web online server (http://glycam.org) to add glycans (biantennary LacNAc N-glycans) on 13 aspartic acids and changed the residue name of "ASN" to "NLN" for residues modified by glycans in the PDB files. In the semi-open and the open simulation systems, the initial structures of the S proteins were generated by SWISS-MODEL with SARS-CoV (*21*) structures of 2 upward RBDs (PDB ID: 6NB6) and 3 upward RBDs (PDB ID: 6NB7) taken as templates. Glycans were also added at the same positions.

Each simulation system was solvated by a cubic TIP3P water box. Considering that the protein mainly stretched in the x and y directions, the boundaries of the water box in each simulation system were extended more in the x and y dimensions than in the z dimension (Table S14) to ensure that the protein stayed in the periodic cell during the simulation process. 3 $Na^+$ ions were added to neutralize the simulation systems. The force fields "leaprc.GLYCAM_06j-1" (*62*), "leaprc.protein.ff14SB" (*63*), and "leaprc.water.tip3p" (*64*) in Amber20 were employed to parameterize glycans, the protein and water molecules (as well as ions), respectively. Topology files suffixed with ".prmtop" and coordinate files suffixed with ".inpcrd" were generated by running the program "tleap" in Amber20. The program "cpptraj" was executed to fix the orders of all atoms.
In order to accelerate simulation speed, we used a timestep of 4 fs via hydrogen mass repartitioning, which repartitioned the mass of hydrogen atoms and their attached heavy atoms (*31*). SHAKE was used to remove the stretching degrees of freedom of hydrogen-involved bonds.

### Simulation process

We first performed 5,000 steps of energy minimization to relax steric clashes, including 2,500 steps of steepest descent minimization and 2,500 steps of conjugate gradient minimization. The temperature of each simulation system was set to 300 K, and the pre-equilibration was run in the NVT ensemble for 300 ps. A radius cutoff of 10 Å was used for non-bonded interactions. Harmonic positional constraints of 10 kcal/mol/$Å^2$ were applied during the pre-equilibration to gradually release improper forces in the system. Before the equilibrium simulation, the solvent was first optimized, followed by optimization of the protein side chains, and then restraints were gradually removed to allow the system to approach a low-energy state. Then we performed 10 ns equilibrium simulation at 300 K for each system. The minimization, pre-equilibration and equilibrium simulation were performed with the binary file "pmemd.cuda" in Amber20.

By adding bias potential to the original potential, aMD (*25*) can promote the protein to overcome local energy barriers, which would enhance the efficiency of conformational sampling. We applied energy boosting to both the whole potential and the torsion term (iamd=3). The aMD parameters adopted in our simulations included the average energy thresholds and inverse strength boost factors for the total potential and the dihedral potential. When estimating these parameters, we scaled the total number of atoms by 0.16 (*65*) in the partially open, semi-open, and open systems but by 0.20 in the closed system, since the last system was more stable. 1000-ns aMD was run for each simulation system with 5 repeats.

We performed tMD to detect the mode characterized by NTD moving out and RBD tilting downward in the partially open system. We calculated the distance between the NTD and the central axis of the S protein in the trajectory of the semi-open system and chose the frame with the largest distance as the target structure. The initial structure was chosen as the last frame of 100 ns equilibrium simulation on the partially open system. tMD (*66*) applies an additional term to the energy function based on the mass-weighted root mean square deviation (RMSD) of a set of atoms in the current structure compared to the reference structure:

$$E = 0.5 \times TGTMDFRC \times NATTGTRMS \times (CURRMSD - TGTRMSD)^2, \quad \text{(Eq. 1)}$$

where TGTMDFRC stands for the force constant, NATTGTRMS stands for the number of Cα atoms of NTD, CURRMSD stands for the RMSD of the current structure compared with the target structure and TGTRMSD stands for expected RMSD value, which is equal to 0 in our work.

### Evaluation of the relative motions between NTD and RBD

To conduct a quantitative analysis of relative motions between the NTD and the RBD, we first defined four stable regions by calculating the residue-wise positional fluctuations (also referred to as root-mean-square fluctuations or RMSF),

$$RMSF_i = \sqrt{\langle (x_i - \langle x_i \rangle)^2 \rangle}, \quad \text{(Eq.2)}$$

where $x_i$ stands for the Cartesian coordinates of the Cα atom of a residue in each frame of the whole trajectory and $\langle x_i \rangle$ stands for the mean value along the simulation trajectory. By selecting residues with small RMSFs, the four stable regions were defined as follows: 1) the RBD core region including resides 334-527, with residues 454-491 excluded; 2) the rotation axis of RBD including residues 541-586; 3) the upper part of the central axis including residues 976-996; and 4) the NTD core region including resides 27-303. We calculated the centroids of each region and defined three evaluation metrics: 1) RBD angle ($\theta_r$) is the scalar angle between centroids of the RBD core region, the rotation axis and the central axis; 2) RBD distance ($d_r$) describes the distance between the RBD core region and the central axis; 3) NTD distance ($d_n$) describes the distance between the NTD core region and the central axis.

We used tICA (*34, 35*) to evaluate motions of the upward RBD in the partially open system when the NTD was forced out by tMD. Instead of using structural similarity as a proxy for kinetic similarity, tICA sought to encode kinetic similarity in the states explicitly by choosing the states along the slowest degrees of freedom. Given a multidimensional dataset, tICA is able to find the linear combinations (tIC) of the input coordinates that maximize the autocorrelation function of the corresponding projection, while restraining this linear combination to be uncorrelated to the previous ones. In practice, we performed tICA on the first 200 ns of the aMD trajectory with a lag time of 2 ns.

## Markov State Model

By using MSMBuilder v3.8.0 (*39, 40*), we built the MSM for the partially open system with the following procedures:

1) To remove the degrees of freedom of rotation and translation, we aligned all frames of all trajectories to the first frame of trajectory 1.
2) As our study focused on relative motions of the NTD and its partner RBD of the S1 region, the Cartesian coordinates of Cα atoms of S1 regions were extracted from five trajectories as input data for MSM.
3) We performed PCA to reduce the dimensions of the input data and projected the input data on the top 4 PCs.
4) The choice of lag-time is important since it allows the discrete microstates to be Markovian and ensures kinetic accessibility (*39, 40*). According to the implied timescale analysis(*41*), if the implied timescale no longer increases at any lag-time longer than $\tau$, then $\tau$ is chosen as a proper lag-time to resolve the process (*37*). We chose the lag time based on MSMs built with a series of microstate numbers including 50, 100, 250, 500, 1000 and 2000. At the beginning, the implied timescale rose gradually, with the 0.2-ns stepwise increment of lag-time increasing. When the lag time reached 1.0 ns, the implied timescales tended to be relatively constant, and no significant increase was detected with any longer lag times in all MSMs with different numbers of microstates. Therefore, we chose 1.0 ns as the lag time.
5) We used the generalized matrix Rayleigh quotient (GMRQ) as a metric to choose a proper number of microstates. GMRQ is a criterion that evaluates how well the MSM eigenvectors generated on the training dataset explains kinetics variance in the test dataset (*42*). We built MSM with a lag time of 1.0 ns and with varying numbers of microstates (50, 100, 250, 500, 1000 and 2000), and finally chose 100 microstates because of the highest value of GMRQ in 5-fold cross validation. With all hyperparameters fixed, all subsequent MSMs were constructed with a lag time of 1.0 ns and microstate number of 100, using mini-batch K-Means (*40*).

6) 100 microstates were lumped into 6 macrostates by perron-cluster cluster analysis (PCCA) (*67*).
7) For the partially open system, MSM was built with 6 macrostates and transition probabilities were calculated between each pair of macrostates.
8) We selected the structures closest to the center of each macrostate as the representative structures.

We also built MSM for the mode characterized by the NTD moving out and the RBD tilting downwards in the semi-open system with the same procedures. 100 microstates were clustered in a 5-dimensional PC space and a lag time of 1.0 ns was chosen for this model. 4 macrostates were selected in the final model.

## Free energy decomposition

We used MM/GBSA for free energy decomposition and detected key residue pairs on the NTD-RBD interface. For each macrostate, we first picked up 100 structures that had the smallest RMSD with the representative structure and discarded structures with RMSD larger than 4 Å. MM/GBSA (*68*) was performed in Amber20 on each microstate for the selected structures. The program "Parmed" was used to split the topology file of the original system into two parts, i.e., a chain with the selected NTD acting as the ligand and the remaining two chains acting as the receptor. We decomposed the total free energy on pairs of residues and identified residue pairs in the NTD of the ligand and in the corresponding RBD of the receptor according to decomposed binding free energies. After ranking the pairwise decomposed binding free energy, we found that most of the residue pairs with high bind affinities were located on the NTD-RBD interface.

## Virtual screening by docking

Virtual screening was performed by AutoDock Vina (*55*, *56*) with the following procedures:

1) We generated the receptor file for docking. The structure of the last frame of the equilibrium simulation in the partially open system was selected as the receptor. Charges were added for each residue.
2) The virtual screening was performed against two compound databases. We culled 9,800 chemical compounds from the category of "in-trails" of ZINC15(*57*-*59*) and 14,249 compounds from the database of TCMSP (*60*). The compounds culled from ZINC15 were approved drugs and investigational compounds in clinical trials with the "sdf" format. The compounds culled from TCMSP were chemical compounds extracted from 502 kinds of Chinese herbs with the "mol2" format. Chemical compounds were converted into the PDB format by OpenBabel (*69*) and were further optimized using the Generalized Amber Force Field (gaff). Charges were added and rotatable bonds were assigned for each compound.
3) We then designed the grid box and set the configuration file. We chose the NTD-RBD interface of the partially open system as the target for docking. We calculated the Cartesian coordinates of the centroid of the NTD and placed this NTD into a rectangular box with each dimension extended by 5Å.
4) We ran AutoDock Vina (*55*, *56*) to screen compounds in two databases with the same configuration. The top 20 compounds with the highest binding affinities in ZINC15 were selected, and the detailed interactions were visualized using Discovery Studio Visualizer. For compounds in TCMSP, we selected all compounds with binding free energy lower than

-9.5 kal/mol. We then analyzed the selected compounds that coexisted in 10 prescriptions recommended by National Health Commission of China for the treatment of COVID-19.

**Acknowledgement:**

We thank Zhuoxuan Liu at the Department of Electrical Engineering of Tsinghua University for his support for drawing the schematic diagram for the NTD “wedge” effect. This work was partially founded by the National Natural Science Foundation of China (no. 31670723, no. 81861138009, and no. 91746119) to H.G. and by the Beijing Advanced Innovation Center for Structural Biology to H.G.

**Author contributions:**

T.W., H.G. and B.S. contributed to the overall methodology and experimental design. J.Z, T.W and S.L. contributed to simulation system design and initial analysis. Y.L., J.Z. and T.W. contributed to intermediate conformations and transition analysis. Y.L., T.W. and Y.W. contributed to interface analysis. T.W. and Y.W. contributed to preliminary drug virtual screening. Y.L., T.W. and J.Z. contributed to writing the original draft. H.G., B.S., T.W. and T.L. contributed to writing. All authors reviewed the final manuscript.

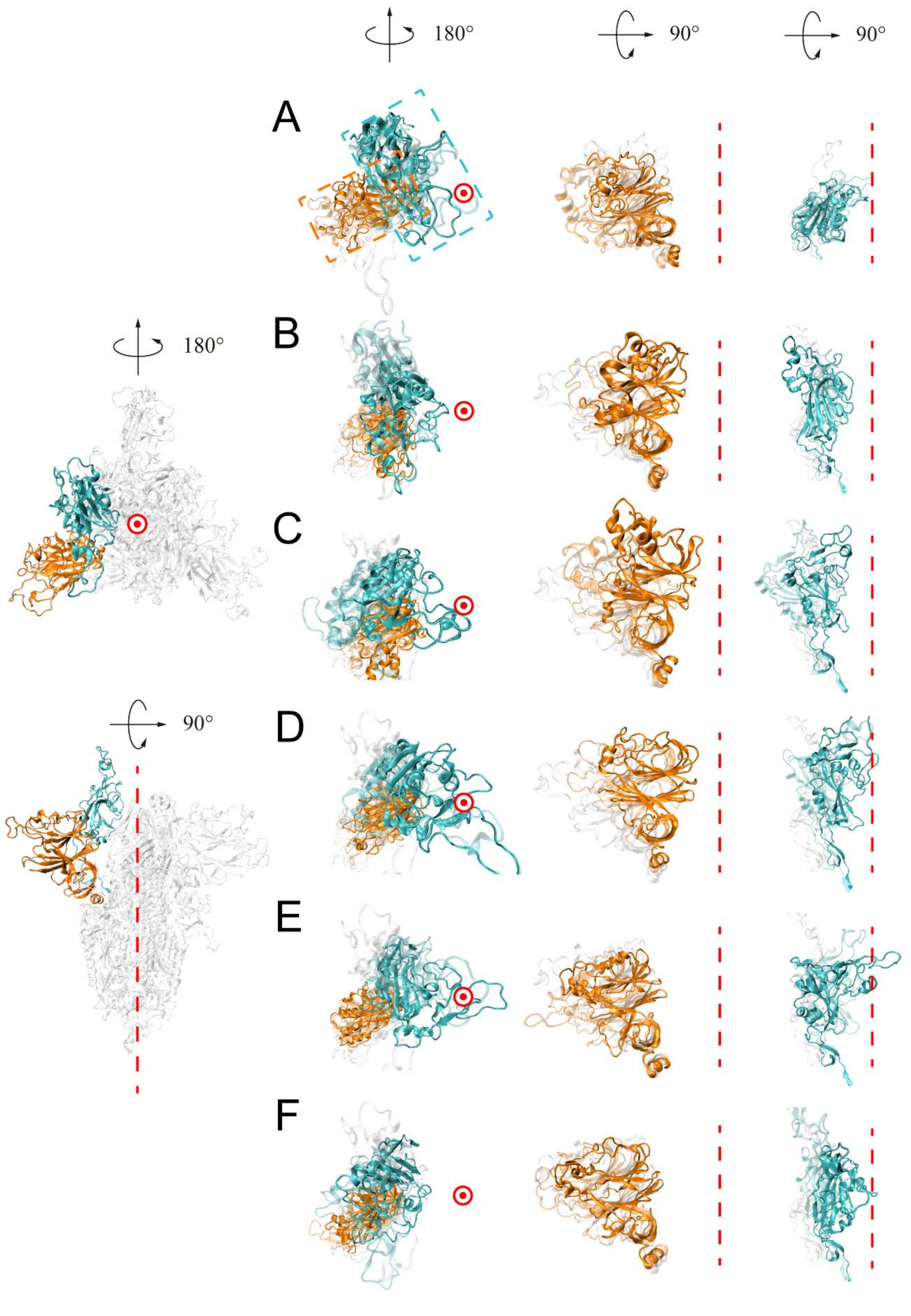

**Fig. 1.** Top view (left) and side view (right) of the initial and final structures of the four simulation systems. The leftmost column is a structural illustration of the pair of RBD and NTD

that is explored in all simulations. Here, the partially open S protein is chosen for demonstration, with the RBD of chain B and the NTD of chain C highlighted in cyan and orange respectively, while the other parts are colored gray in the transparent mode. The central axis is represented by a red dashed line in the side view and a red circle in the top view. (A-F) Comparison of the initial and final structures in all simulation systems. In the second column, the RBD of interest is colored cyan and its partner NTD is colored orange. The initial structures are shown in the transparent mode to assist visual identification of conformational changes. The NTD and RBD are enlarged and shown in the following two columns with the side view. The initial structures are colored gray in the transparent mode. In the final structure, the RBDs are colored cyan and the NTDs are colored orange. The intermediate states are shown in the transparent mode with the RBDs colored cyan and the NTDs colored orange. In the closed system (A), all RBDs are locked in the downward orientation. In the partially open system (B), semi-open system (C) and open system (D), the tendency of the upward RBD to tilt downwards is prohibited by the neighboring NTD moving toward the central axis. (E-F) In the semi-open system (E) and the open system (F), the upward RBD reorients to the downward orientation when the NTD swings away from the central axis and then moves back.

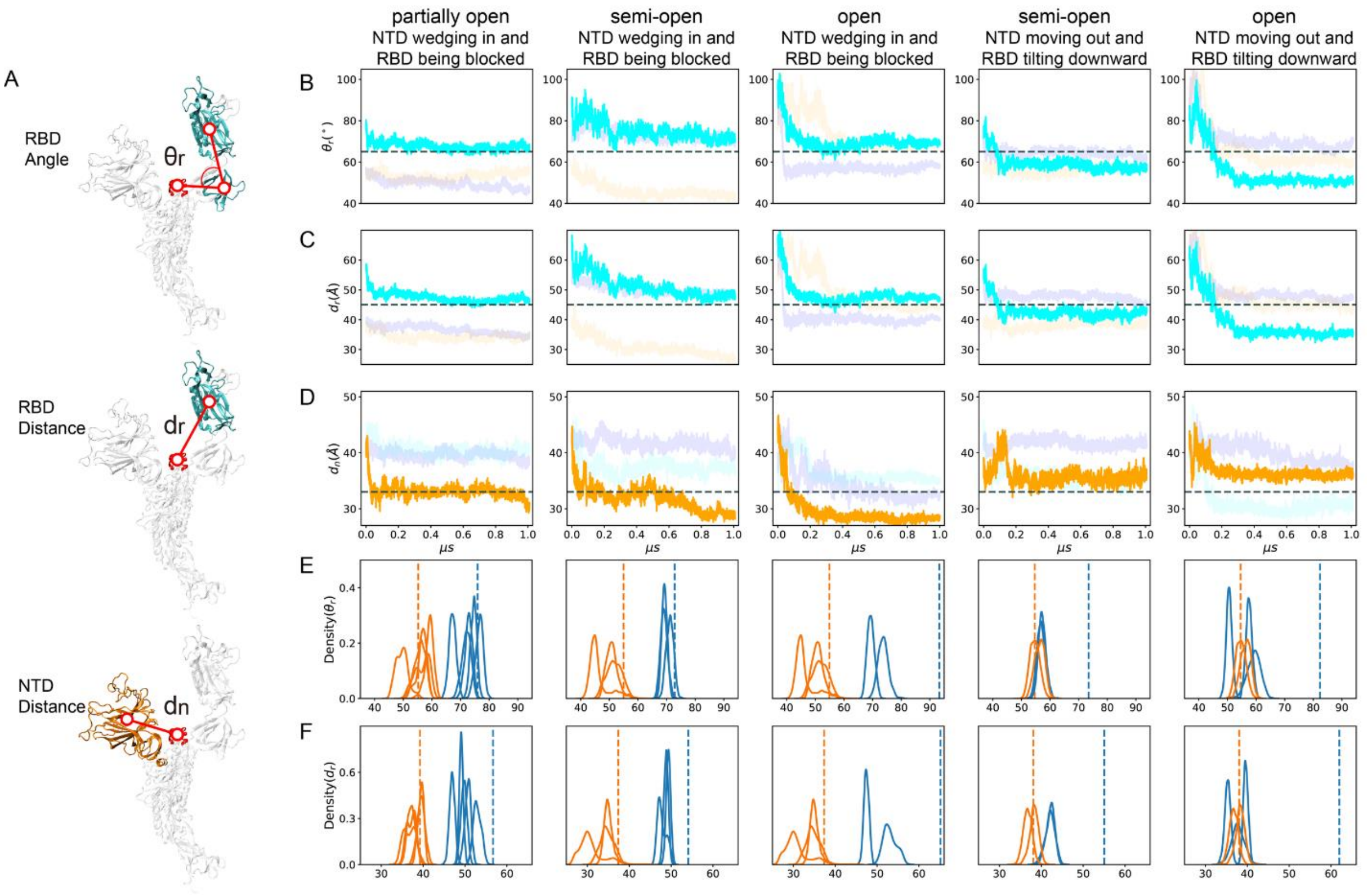

**Fig. 2.** Illustration of the relative motions between NTD and RBD in the four simulation systems. (A) The sketch of three quantitative metrics: the angle (θr) and distance (dr) to quantify RBD movement and the distance (dn) to evaluate NTD movement. (B-D) Evolution of the three metrics in the whole simulation. For each simulation system, an upward RBD and its adjacent NTD are picked from one repeat for evaluation. Chains with the selected NTD and RBD are colored orange and cyan, respectively, whereas the remaining chains are shown in the transparent mode. The dashed lines are shown as reference to capture the differences between simulation systems. (E-F) Probability distributions of the angle (θr) and the distance (dr) as collected from the simulations to illustrate the movement of the upward RBDs of interest (blue solid lines). The distributions of the RBDs held at the downward orientations (orange solid lines) are taken as control. The blue dashed lines and the orange dashed lines indicate the corresponding values of the initial structures of the upward and downward RBDs, respectively. For each simulation system, one chain from each of the 5 repeats is collected for evaluation. To evaluate the final state of the upward RBDs during simulation, the distributions of the upward RBDs are estimated based on the last 200 ns trajectory of each repeat, while the control distributions of the downward RBDs (chain A in both partially open and semi-open systems) are calculated from the whole trajectories. The first three columns in (B-F) show the blockage of the RBD by its adjacent NTD in the partially open, semi-open and open systems (corresponding to Fig. 1B-D), respectively, while the last two columns show the complete downwards tilt of the RBD in the semi-open and open systems (corresponding to Fig. 1E&F). Considering that the open system lacks downward RBDs, in (E) and (F), the downward RBDs of the semi-open system (corresponding to Fig. 1B) are taken as the control.

A

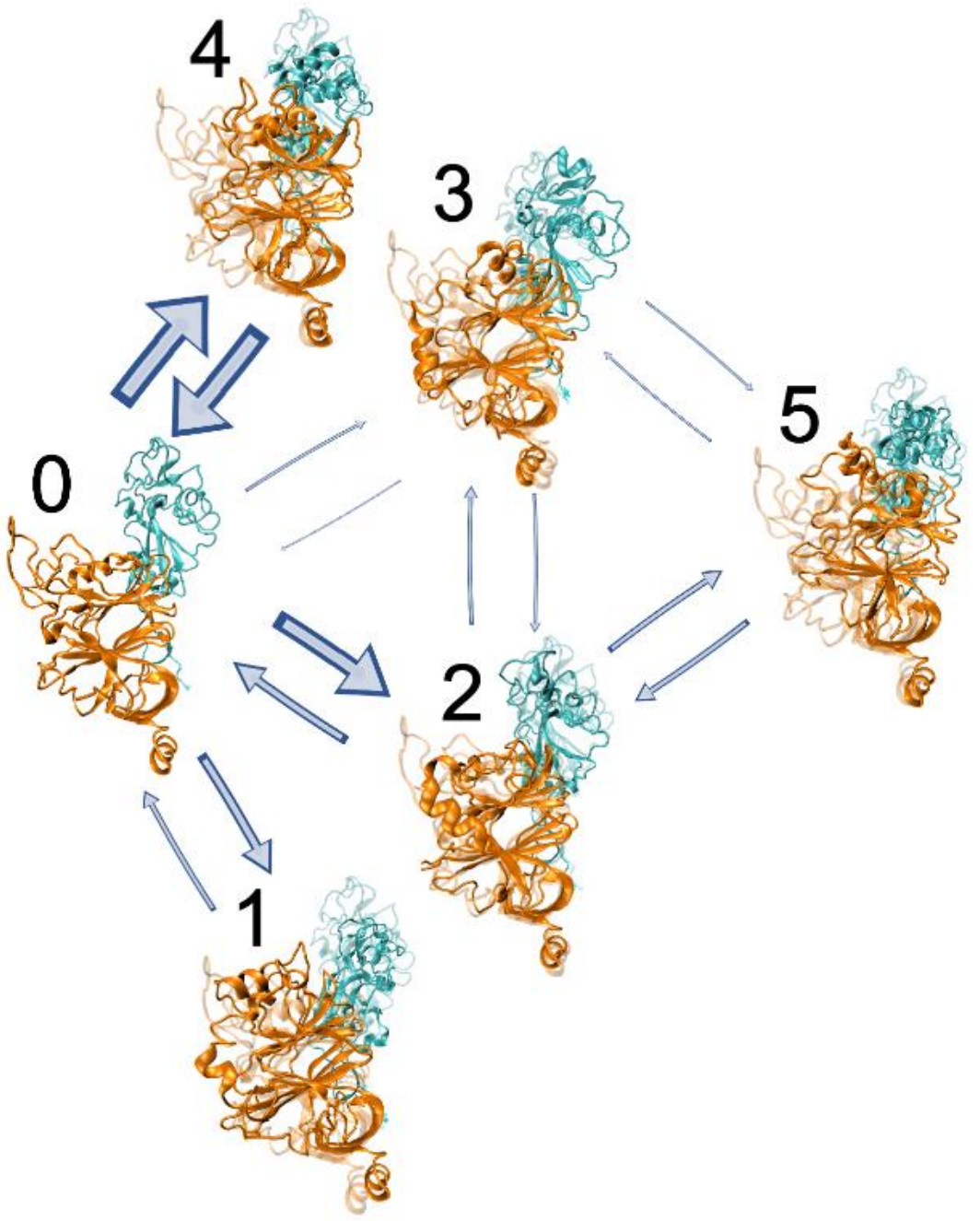

B

| State | State ID | $d_n$ | $\theta_r$ | $d_r$ |
| --- | --- | --- | --- | --- |
| Initial | 0 | 34.4 ± 3.1 | 73.3 ± 3.0 | 52.3 ± 1.6 |
| Intermediate | 1 | 35.1 ± 1.7 | 75.2 ± 1.7 | 49.4 ± 1.8 |
| Intermediate | 2 | 35.2 ± 1.8 | 70.7 ± 2.8 | 52.6 ± 1.7 |
| Intermediate | 3 | 33.0 ± 0.8 | 67.5 ± 1.6 | 47.2 ± 1.2 |
| Final | 4 | 31.4 ± 0.6 | 73.5 ± 1.3 | 50.3 ± 0.8 |
| Final | 5 | 30.8 ± 1.2 | 74.8 ± 2.1 | 51.5 ± 1.2 |

C

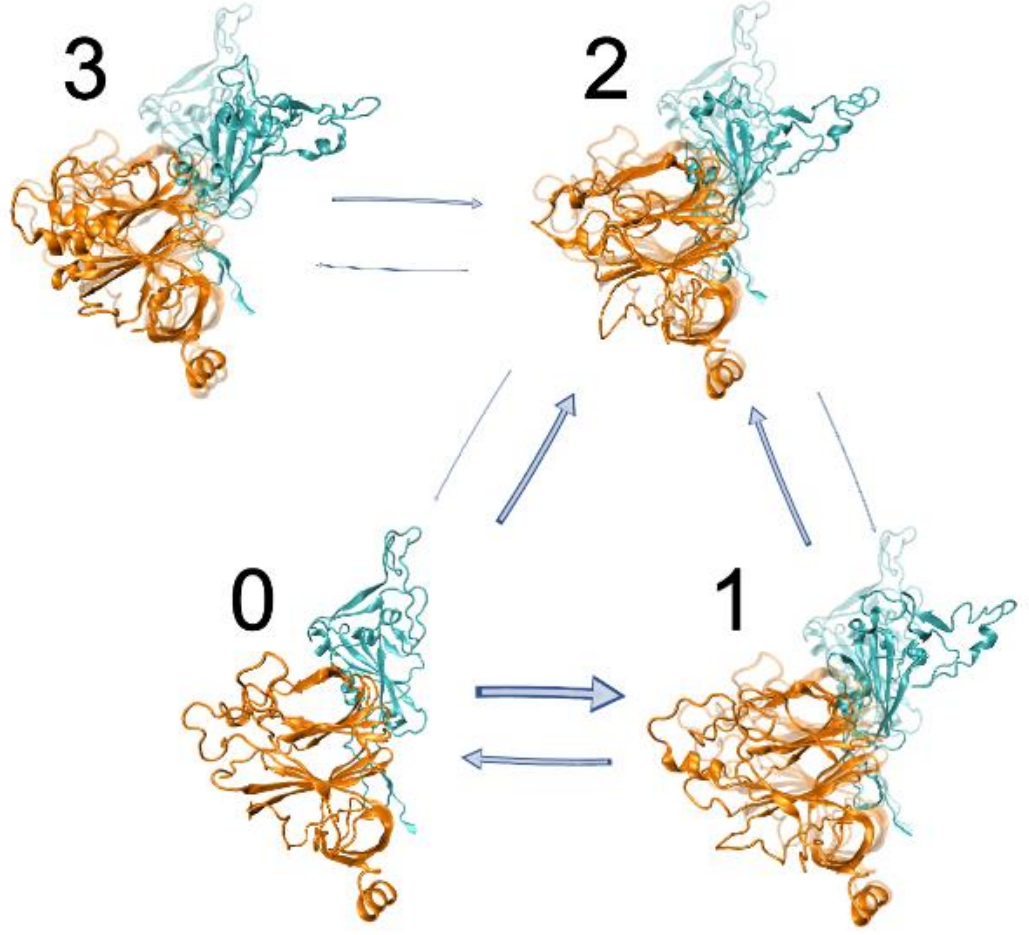

D

| State | State ID | $d_n$ | $\theta_r$ | $d_r$ |
| --- | --- | --- | --- | --- |
| Initial | 0 | 36.5 ± 1.6 | 72.6 ± 4.7 | 52.2 ± 3.8 |
| Intermediate | 1 | 37.8 ± 1.3 | 65.6 ± 7.1 | 47.2 ± 3.8 |
| Intermediate | 2 | 36.5 ± 2.3 | 58.9 ± 2.0 | 42.7 ± 1.3 |
| Final | 3 | 35.4 ± 0.8 | 57.2 ± 1.4 | 41.8 ± 1.1 |

**Fig. 3.** MSMs on the mode characterized by the NTD wedging in and the RBD being blocked in the partially open system and the mode characterized by the NTD moving out and the RBD tilting down in the semi-open system.

(A) MSM in the partially open system illustrates the mode characterized by NTD wedging in and RBD being blocked. Six macrostates were extracted from 5 trajectories, among which state 0 is the initial state with an upward RBD, states 1, 2 and 3 are intermediate states, while states 4 and 5 are two final states. All structures are superposed onto the initial structures (shown in the transparent mode) to facilitate visual detection of structural changes. The thickness of arrows is proportional to the transition probability. Bidirectional arrows indicate mutual transition. (B) A quantitative evaluation on all macrostates in (A). Final states have smaller NTD distance (dn) but comparable RBD angle (θr) and distance (dr) when compared with the initial state. (C) MSM in the semi-open system illustrates the mode characterized by NTD moving out and RBD tilting downward. Four macrostates were extracted, among which state 0 is the initial state with an upward RBD, states 1 and 2 are intermediate states, while state 3 is the final state. All structures are superposed onto the initial structures (shown in the transparent mode) to facilitate visual detection of structural changes. (D) A quantitative evaluation on all macrostates in (C). Both RBD angle (θr) and distance (dr) in the final state are remarkably smaller than those in the initial state.

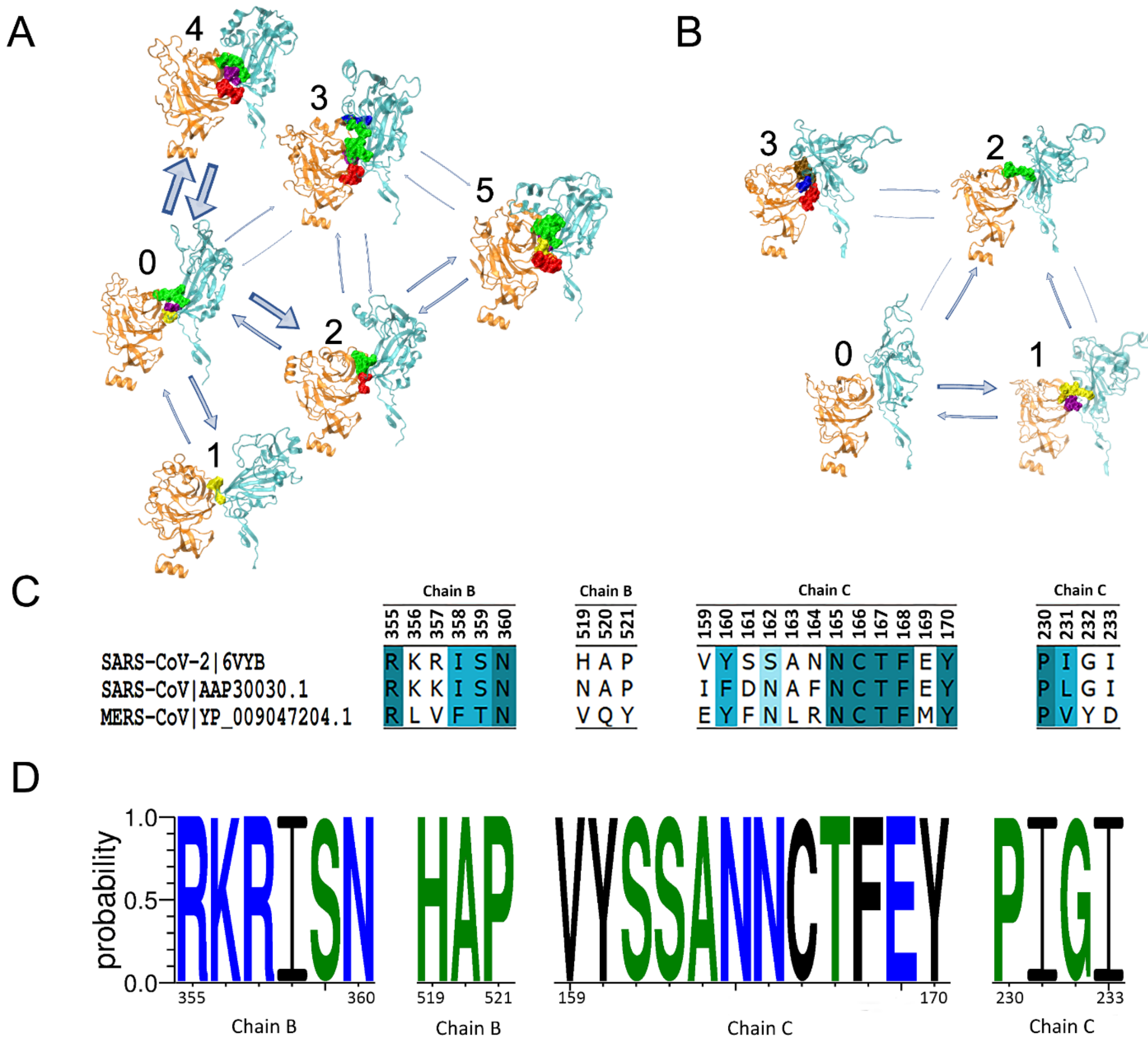

**Fig. 4.** Key residue pairs identified at the interacting interface between the NTD and corresponding RBD during the transition process.

(A) Key residue pairs of each macrostate in MSM for the mode characterized by NTD wedging in and RBD being blocked. Spatially neighboring residue pairs form a clump and different clumps are marked with different colors (see Table S2 for more details). (B) Key residue pairs of each macrostate in MSM for the mode characterized by NTD moving out and RBD tilting downward (see Table S3 for more details). (C) Evolutional analysis of key residue pairs in (A) among SARS-CoV-2 (PDB ID:6VYB) (9), SARS-CoV (GenBank: AY278488.2) (48) and MERS-CoV (GenBank: NC_019843.3) (49). Results were visualized by Discovery Studio Visualizer (50). (D) Evolutional analysis of key residue pairs in (A) among all samples of COVID-19 patients from GISAID (51, 52). Data were collected on September 28th, 2020. Results were visualized by WebLogo (53, 54). Hydrophilic, neutral and hydrophobic residues are colored in blue, green and black, respectively.

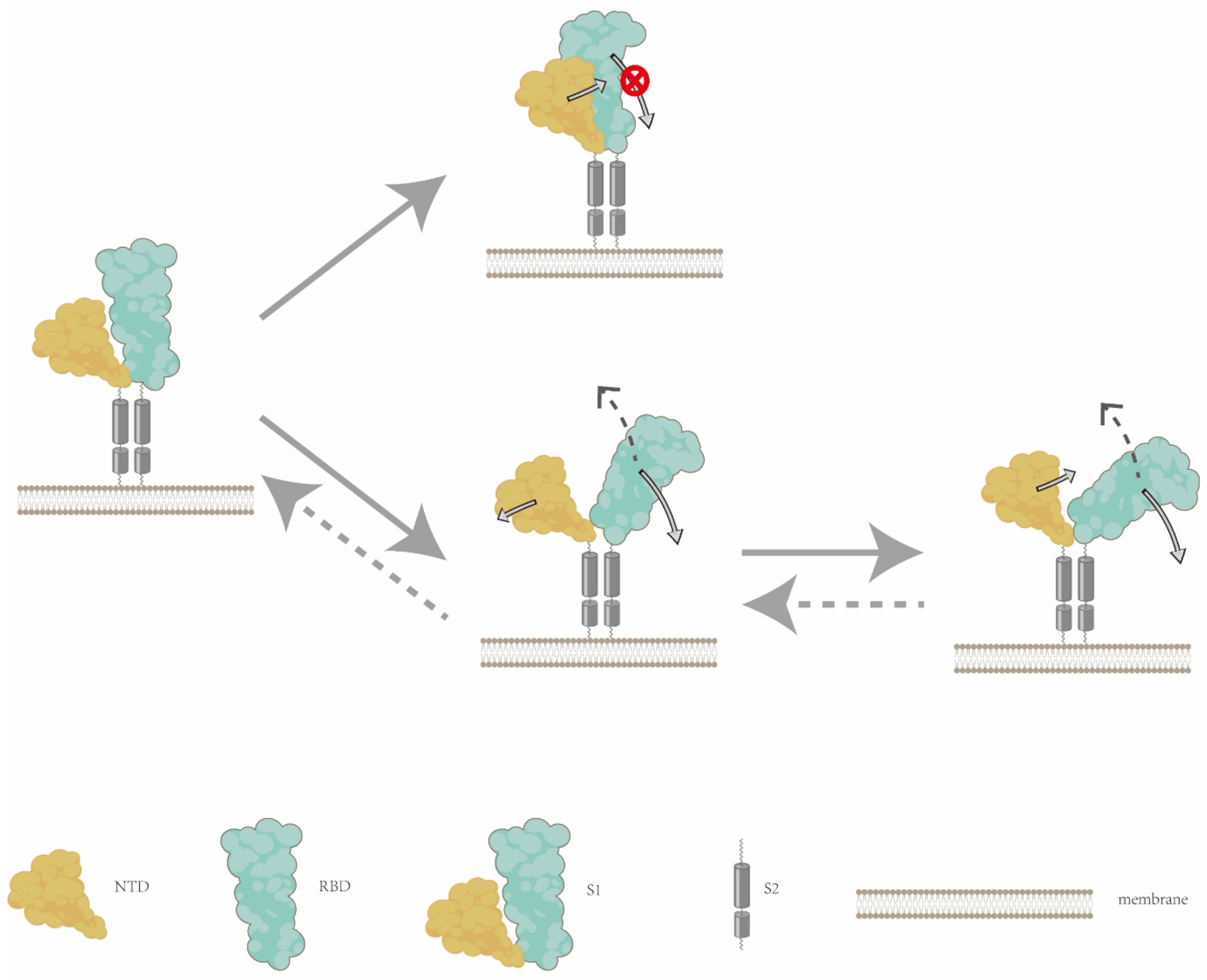

**Fig. 5.** The schematic diagram for the NTD “wedge” effect in controlling conformational changes of the S protein.

The leftmost structure is the simplified initial conformation, where only one NTD (orange) and its neighboring RBD (cyan) are showed in this sketch. When the NTD wedges in and favorably interacts with the RBD, the free tilting of the RBD is blocked and the S protein is locked into individual conformation states. Occasionally, when the NTD swings out and detaches from the RBD, the RBD’s reorientation is restored, thus enabling the inter-state transition of the S protein. Hypothetically, the opposite motion of the RBD tilting upward may occur during the activation of the S protein, which is illustrated by dotted arrows.

A

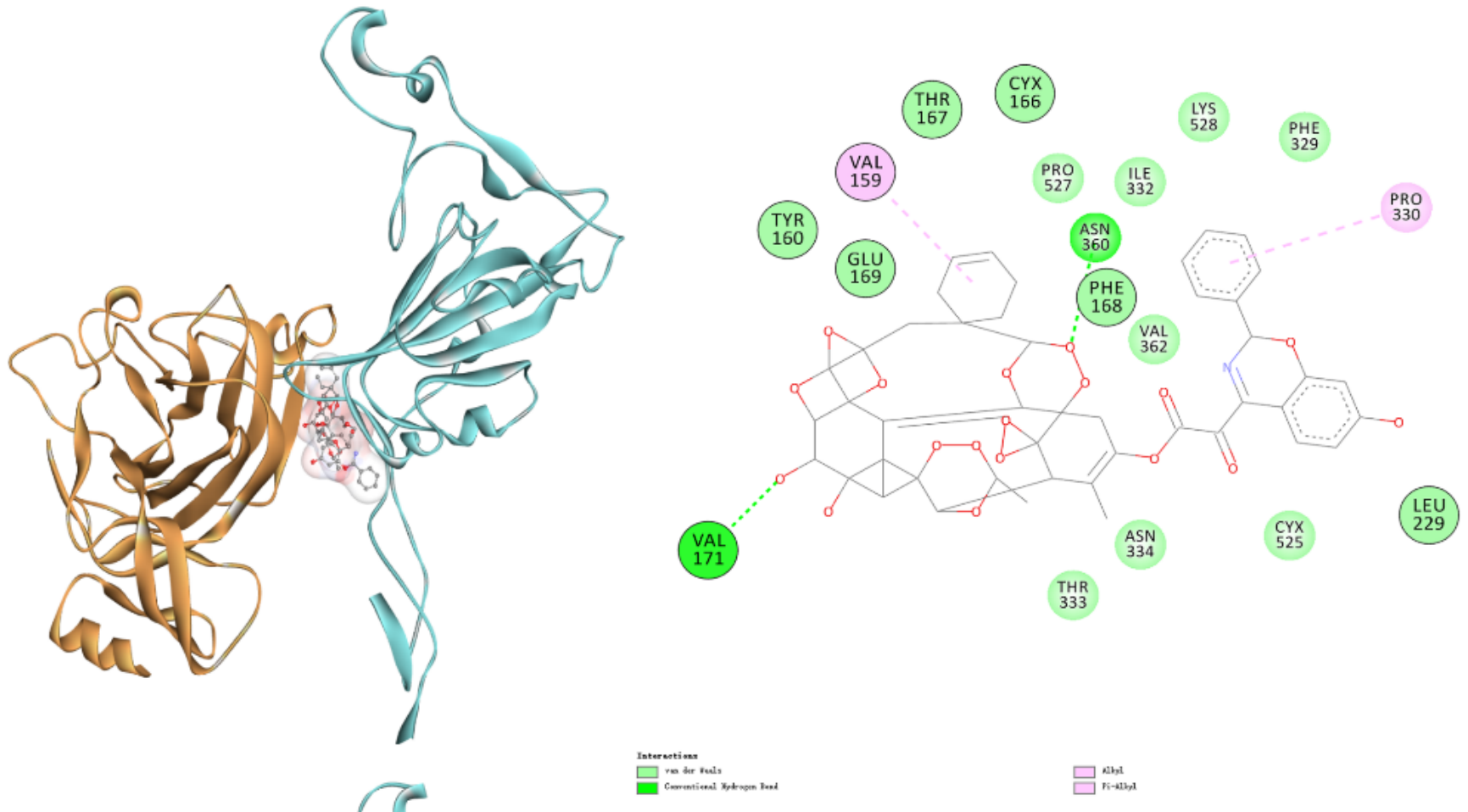

B

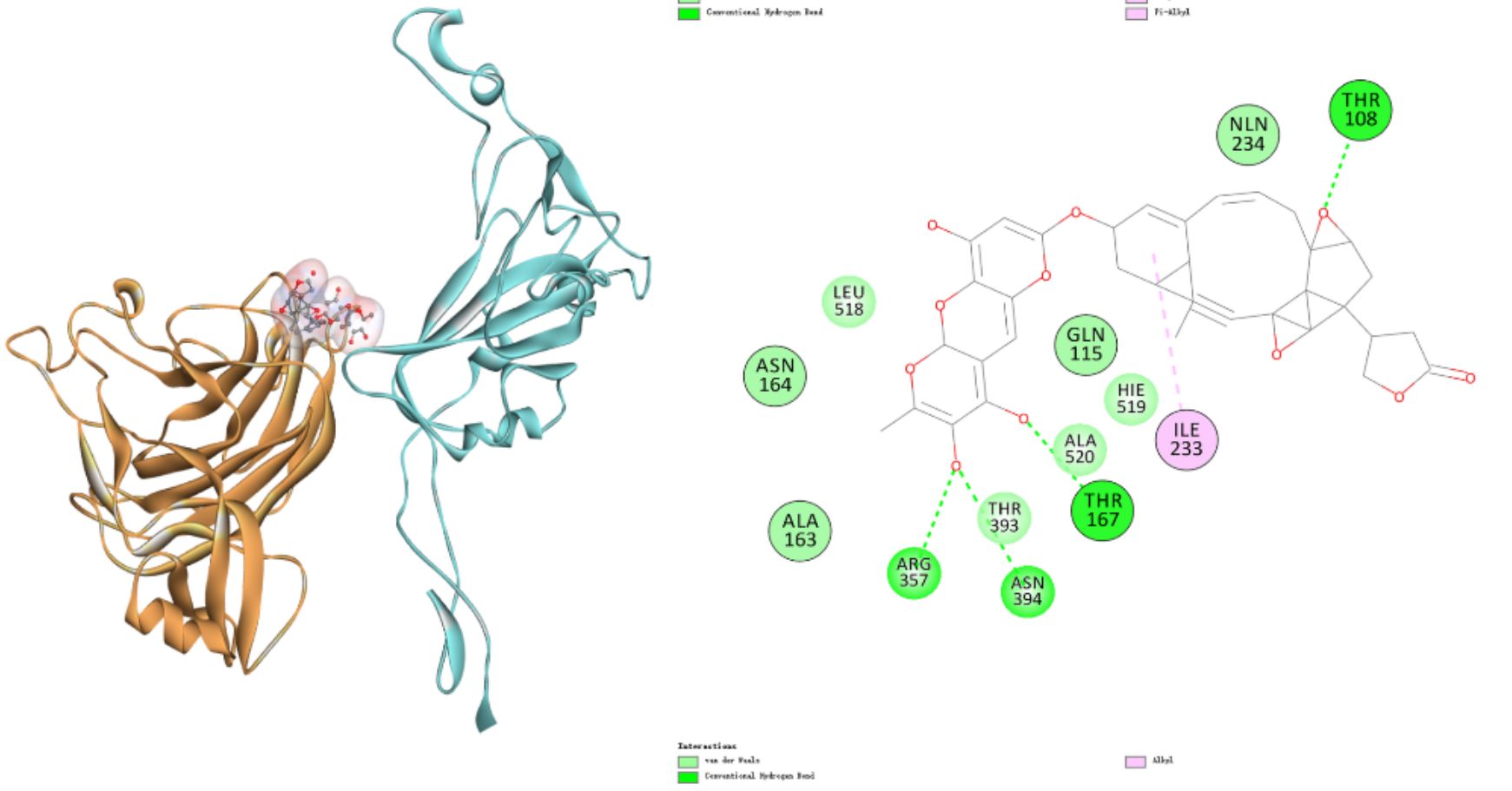

C

| Herb name | Detected compounds | Binding Affinity |
|---|---|---|
| Polyporus Umbellatus(Pers)Fr. | polyporoid,b | -9.5 |
| Atractylodes Macrocephala Koidz. | biatractylolide | -9.5 |
| Radix Bupleuri | 3'-O-Acetylsaikosaponin D_qt | -9.8 |
| | Saikosaponin D | -9.7 |
| | saikosaponin c | -9.5 |
| | saikosaponin c_qt | -9.5 |
| Scutellariae Radix | 5-o-caffeoylquinic acid | -9.8 |
| Asteris Radix Et Rhizoma | aster saponin B | -10.3 |
| | echinocystic acid | -9.7 |
| | kaempferol | -9.7 |
| | (4aR,5R,6aR,6aS,6bR,8aR,10R,11S,12aR,14bS)-5,10,11-trihydroxy-2,2,6a,6b,9,9,12a-heptamethyl-1,3,4,5,6,6a,7,8,8a,10,11,12,13,14b-tetradecahydropicene-4a-carboxylic acid | -9.6 |
| | aster saponin E | -9.6 |
| Rhizoma Dioscoreae | diosgenin | -9.5 |
| Aurantii Fructus Immaturus | isosakuranetin-7-rutinoside | -9.6 |
| | naringenin-4'-glucoside-7-rutinoside | -9.5 |
| | (2S)-7-[(2S,3R,4S,5S,6R)-4,5-dihydroxy-6-methylol-3-[(2S,3R,4R,5R,6S)-3,4,5-trihydroxy-6-methyl-tetrahydropyran-2-yl]oxy-tetrahydropyran-2-yl]oxy-5-hydroxy-2-(3-hydroxy-5-methoxy-phenyl)chroman-4-one | -9.5 |
| | Scolymoside | -9.5 |
| Citrus Reticulata | (2S)-7-[(2S,3R,4S,5S,6R)-4,5-dihydroxy-6-methylol-3-[(2S,3R,4R,5R,6S)-3,4,5-trihydroxy-6-methyl-tetrahydropyran-2-yl]oxy-tetrahydropyran-2-yl]oxy-5-hydroxy-2-(3-hydroxy-5-methoxy-phenyl)chroman-4-one | -9.5 |

**Fig. 6.** Case study of virtual screening by docking.

(A-B) top compounds, ZINC000261494659 (A) and ZINC000256109538 (B) from ZINC15 database are identified with high binding affinities to the NTD-RBD interface in the partially open system. In the left column, the NTD of chain C is colored orange and the upward RBD of chain B is colored cyan. The drug compounds are represented with the ball-and-stick model. In the right column, detailed interactions between the compounds and the NTD-RBD interface are visualized using Discovery Studio Visualizer. Residues in chain C are highlighted with a black circle. (C) Compounds from the TCMSP database with high binding affinities are detected in the traditional Chinese prescription “Qing Fei Pai Du Tang”.